\documentclass[11pt]{article} 
\usepackage{longtable,amsmath,amssymb}
\usepackage{array} 
\usepackage{float}
\usepackage[utf8]{inputenc} 
\usepackage{geometry} 
\usepackage{array}
\usepackage{amsmath}
\usepackage{amsthm}
\usepackage{graphicx} 
\usepackage{longtable}
\usepackage{multirow}
\usepackage{multicol}

\usepackage{array}

\usepackage{booktabs} 
\usepackage{array} 
\usepackage{paralist} 
\usepackage{verbatim} 
\usepackage{subfig} 
\usepackage{color}

\usepackage{fancyhdr} 
\usepackage{sectsty}
\allsectionsfont{\sffamily\mdseries\upshape} 

\usepackage[nottoc,notlof,notlot]{tocbibind} 
\usepackage[titles,subfigure]{tocloft} 

\title{The energy-momentum tensor in relativistic kinetic theory: the role of the center of mass velocity in the transport equations for multicomponent mixtures}
\author{A. R. Sagaceta-Mejía, A. Sandoval-Villalbazo, \\  J.H. Mondragón-Suárez  \\
	Departamento de Física y Matemáticas,  Universidad Iberoamericana \\
	Prolongación Paseo de la Reforma 880, Lomas de Santa Fe 01219. \\
	Mexico City, Mexico}

\begin{document}
	\maketitle 
	\begin{abstract}
Relativistic kinetic theory is applied to the study of  the balance equations for relativistic multicomponent mixtures, comparing the approaches corresponding to  Eckart's and Landau-Lifshitz's frames. It is shown that the  concept of particle velocity relative to the center of mass  of the fluid is essential to establish the structure of the energy-momentum tensor in both cases. Different operational definitions of the center of mass velocity lead either to the inclusion of heat in the energy-momentum tensor (particle/Eckart frame) or to strictly relativistic contributions to the diffusion fluxes (energy/Landau-Lifshitz frame).
	The results here obtained are discussed emphasizing the physical features regarding each approach.
\end{abstract}



\section{Introduction}

The transport equations for dissipative relativistic mixtures are not completely
understood. In particular, the  precise form of the relations  between thermodynamic forces and fluxes has been  a subject of debate for decades.  Purely relativistic features of high temperature fluids in non-equilibrium states were first identified in the pioneering work by C. Eckart \cite{eckart1940thermodynamics} and later discussed in the classical textbook written by L. Landau and E.M. Lifshitz \cite{Landau}. Moreover,  relativistic kinetic theory have provided insight to the microscopic foundations of the transport equations, allowing to the identification of open problems \cite{Israel,struchtrup,cercignani2002relativistic,garcia2012microscopic}.  Once the generic instabilities of  relativistic simple fluids were phenomenologically  identified by Hiscock and Lindlom back in 1985 \cite{Hiscock}, the corresponding analysis of the physics behind them became a subject of study by several authors \cite{cercignani2002relativistic, PeterVan, Koide, Smerlak,  SandovalVillalbazoPhisicaA}. The
stability properties of relativistic fluids are also a subject of	debate. Although first order in the gradients theories have been improved	through the use of kinetic theory \cite{SandovalVillalbazoPhisicaA} other issues regarding	stability and causality had motivated the use of second order theories	and generalized formalisms in order to establish proper sets of transport 	equations \cite{muller, israelextended,stewart,Tsumura}. In this context, the use of the Chapman-Enskog treatment of the relativistic Boltzmann equation  became a particularly useful tool in order to  examine the effect of the constitutive equations on the stability of linear fluctuations in Eckart's frame \cite{GARCIAPInest}.

The establishment of the stress tensor in non-relativistic kinetic theory involves the decomposition of the velocity field of single particles in terms of the hydrodynamic velocity and a ``peculiar" or ``chaotic" velocity measured in a system comoving with the center mass of the fluid. This type of decomposition is also valid for the case of multicomponent systems, and eventually leads to the  expressions for diffusive fluxes, heat flows and viscous dissipation. One possible approach for the analysis of relativistic multicomponent systems is based on the use of a similar decomposition that is equivalent to the Lorentz transformation.

The purpose of this paper is to show that the structure of the energy-momentum tensor for a multicomponent relativistic fluid, as established from kinetic theory, critically depends on the operational definition of the center of mass velocity of the system. The usual expression  of the energy-momentum tensor in Eckart's frame, which includes the heat flux, is obtained by means of a ``non-thermal" definition of the center of mass velocity. In contrast, in the Landau-Lifshitz approach  the heat flux is not contained in the energy-momentum tensor and new dissipative effects appear in the diffusive fluxes established through the corresponding center of mass velocity definition, which now includes thermal energy.

This paper is divided as follows: in section two we review the non-relativistic establishment of the stress tensor for a multicomponent fluid emphasizing how the definition of the center of mass velocity is motivated by the requirement that the cross-terms present in the stress tensor vanish. In section three it is shown that the presence of the cross-terms in Eckart's frame is directly related to the heat flux, while in Landau's frame the heat flux will not be present in the energy-momentum tensor if the center of mass velocity is suitable defined.  Final remarks concerning  the analysis of the physical features of both approaches, and its relation with  the generic instabilities of multicomponent relativistic fluids  are included in the last section of this work.

\section{The stress tensor in the non-relativistic regime}
The non-relativistic Boltzmann equation for multicomponent systems in the absence of  external forces  is given by

\begin{equation}
\color{black}
\frac{\partial f_{(i)}}{\partial t}+\vec{v}_{(i)}\cdot\frac{\partial f_{(i)}}{\partial\vec{r}_{(i)}}=
\sum_{j=1}^{{k}} \mathsf{J}(f_{(i)}f_{(j)})
,\quad i=1,2,\ldots,k.
\label{eq:4-Bolt2}
\end{equation}
where   $\vec{v}_{(i)}$ is the velocity vector, $\vec{r}_{(i)}$ the position of the particle,    $f_{(i)}$ is the distribution function  for the  $i-th$ species and  $\mathsf{J}(f_{(j)}f_{(i)})$ is the   collision kernel   \cite{Chapman1970}. The balance equations for particles and linear momentum are obtained  multiplying Eq. (\ref{eq:4-Bolt2}) by the collision invariants $\psi_{(i)}= [m_{(i)},m_{(i)}   \vec{v}_{(i)}]$ respectively,  and integrating with respect to the velocity fields. For the particle balance we use  the collision invariant  $m_{(i)}$ to get,

\begin{equation}
\frac{\partial}{\partial t}\left(m_{(i)} \int f_{(i)}d\vec{v}_{(i)}\right)+\nabla\cdot\left(m_{(i)} \int \vec{v}_{(i)}f_{(i)}d\vec{v}_{(i)}\right)=0.
\label{eq:4-balanceparticulas}
\end{equation}
At this point, we introduce the definition of statistical average of a dynamical variable $\psi_{(i)}$. In the non-relativistic case, this quantity is simply  given by
\begin{equation}
\left\langle \psi_{\left(i\right)}\right\rangle =\frac{1}{n_{(i)}}\int\psi_{\left(i\right)}f_{\left(i\right)}d\vec{v}_{\left(i\right)},
\label{eq:4-promnr}
\end{equation}
In the absence of dissipation (Euler regime), the Maxwellian distribution function is used for the calculation of the statistical averages,  namely
\begin{equation}
f_{(i)}=f_{(i)}^{(0)}=n_{(i)} \left(\frac{m_{(i)}}{2\pi k_B T}\right)^{3/2} \exp\left[{-\frac{m_{(i)}\left(\vec{v}_{(i)}-\vec{u}\right)^2}{2k_B T}}\right].
\label{eq:Maxwell}
\end{equation}
where $k_B$ is the Boltzmann constant and $T$ is the local temperature. Notice that $n_{(i)}$ is the particle number density of the  $i-th$ species and that it is given by
\begin{equation}
n_{(i)}=\int f^{(0)}_{(i)} d\vec{v}_{(i)}.
\end{equation}
The relation between the molecular   velocity $\vec{v}_{(i)}$, the chaotic velocity $\vec{C}_{(i)}$ of a single particle   and the center of mass velocity $\vec{u}$ for the mixture  is given by the Galilean transformation

\begin{equation}
\vec{v}_{\left(i\right)}=\vec{C}_{\left(i\right)}+\vec{u}.
\label{eq:4-vel}
\end{equation}
Eq. (\ref{eq:4-vel}) is expressed in terms of the chaotic velocity of the $i-th$ species  and the  center of mass velocity $\vec{u}$ of the fluid, \textit{which has not been defined yet}.     Here, the hydrodynamic velocity for each species is given by:
\begin{equation}
\vec{u}_{(i)}=\langle\vec{v}_{(i)}\rangle=\frac{1}{n_{(i)}}\int \vec{v}_{(i)} f_{(i)}d\vec{v}_{(i)}, 
\label{eq:4-vel2}
\end{equation}
Following the standard procedure, we use  Eqs.  (\ref{eq:4-vel}) and (\ref{eq:4-vel2}) so that the particle balance equation (\ref{eq:4-balanceparticulas}) can be rewritten as:

\begin{equation}
\frac{\partial}{\partial t}\left( m_{(i)}n_{(i)}\right)+\nabla\cdot\left( m_{(i)} n_{(i)}
\left\langle\vec{C}_{\left(i\right)}
\right\rangle
+m_{(i)} n_{(i)} \vec{u}
\right)=0,
\label{eq:4cont} 
\end{equation}
From Eq. (\ref{eq:4cont}) it is direct to identify the diffusive flux for species $i$ as:

\begin{equation}\vec{\mathcal{J}}_{(i)}= m_{(i)} n_{(i)} \left\langle\vec{C}_{\left(i\right)}
\right\rangle,
\label{eq:difusivo} \end{equation}  
so that the continuity equation can also be expressed as:
\begin{equation}
\frac{\partial}{\partial t}\left( m_{(i)}n_{(i)}\right)+\nabla\cdot\left( 
m_{(i)} n_{(i)} \vec{u}
\right)=-\nabla \cdot\vec{\mathcal{J}}_{(i)}
\label{eq:4bcont} 
\end{equation}
In order to obtain the particle balance equation for the whole mixture, we perform a summation of Eq. (\ref{eq:4bcont}) over all species obtaining:
\begin{equation}\frac{\partial}{\partial t}\left(\sum_{i=1}^k
n_{(i)}m_{(i)}\right)+\nabla\cdot\left[\sum_{i=1}^k\left(n_{(i)}m_{(i)}\right)\vec{u}\right]=- \sum_{i=1}^k \nabla\cdot\vec{\mathcal{J}}_{(i)},
\label{eq:dif2}
\end{equation}
The procedure mentioned above, applied to the collision invariant $m_{(i)} \vec{v}_{(i)}$ leads to the equation of motion:
\begin{equation}
\frac{\partial}{\partial t}\left(m_{(i)} \int \vec{v}_{\left(i\right)}f_{\left(i\right)}d\vec{v}_{\left(i\right)}\right)+\nabla\cdot\left(m_{\left(i\right)} \int \vec{v}_{\left(i\right)}\vec{v}_{\left(i\right)}f_{\left(i\right)}d\vec{v}_{\left(i\right)}\right)=0.\label{eq:4-cons}
\end{equation}
Using the average notation (\ref{eq:4-promnr}) and summing over all species we can rewrite Eq. (\ref{eq:4-cons}) as:

\begin{equation}
\frac{\partial}{\partial t}\left[\sum_{i=1}^{k}m_{\left(i\right)}n_{\left(i\right)}\left\langle \vec{v}_{\left(i\right)}\right\rangle \right]+\nabla\cdot\left[\sum_{i=1}^{k}m_{\left(i\right)}n_{\left(i\right)}\left\langle \vec{v}_{\left(i\right)}\vec{v}_{\left(i\right)}\right\rangle \right]=0
\label{eq:4-grad}
\end{equation}
This equation can be expressed in terms of the chaotic velocities $\vec{C}_{\left(i\right)}$ leading to

\begin{align}
\frac{\partial}{\partial t}\left[\vec{u}\left(\sum_{i=1}^{k}m_{\left(i\right)}n_{\left(i\right)}\right)+\sum_{i=1}^{k}m_{\left(i\right)}n_{\left(i\right)}\left\langle \vec{C}_{\left(i\right)}\right\rangle
\right]
+\nabla\cdot\left[
\sum_{i=1}^{k}m_{\left(i\right)}n_{\left(i\right)}\left\langle \vec{C}_{\left(i\right)}\vec{C}_{\left(i\right)}\right\rangle
\right. \nonumber\\ 
\left.+2\vec{u}\left(\sum_{i=1}^{k}m_{\left(i\right)}n_{\left(i\right)}\left\langle \vec{C}_{\left(i\right)}\right\rangle \right)+
\vec{u}\vec{u}\left(\sum_{i=1}^{k}m_{\left(i\right)}n_{\left(i\right)}\right) \right]& =0. \label{eq:4-gradd}
\end{align}

The terms  of the type $\vec{C}_{\left(i\right)}\vec{u}$ present in Eq. (\ref{eq:4-gradd}) only vanish with a suitable definition of the center of mass velocity $\vec{u}$. Indeed,  the requirement
\begin{equation}
\sum_{i=1}^{k}m_{\left(i\right)}n_{\left(i\right)}\left\langle \vec{C}_{\left(i\right)}\right\rangle=0,
\label{eq:4-sumac}
\end{equation}
can be expressed, using   Eq. (\ref{eq:4-vel}) as

\begin{equation}
\sum_{i=1}^{k}m_{\left(i\right)}n_{\left(i\right)}\vec{u}_{\left(i\right)}-\vec{u}\left(\sum_{i=1}^{k}m_{\left(i\right)}n_{\left(i\right)}\right)=0.
\label{eq:4-sumau}
\end{equation}
Solving Eq. (\ref{eq:4-sumau}) for $\vec{u}$, the center of mass velocity is easily identified as:

\begin{equation}
\vec{u}_{cm}=\frac{\sum_{i=1}^{k}m_{\left(i\right)}n_{\left(i\right)}\vec{u}_{\left(i\right)}}{\sum_{i=1}^{k}m_{\left(i\right)}n_{\left(i\right)}}.
\label{eq:4-barinorel}
\end{equation}
The  definition (\ref{eq:4-barinorel}) simplifies both the balance equations (\ref{eq:dif2}) and (\ref{eq:4-gradd}). Indeed, the right hand side of the mass balance equation (\ref{eq:dif2}) vanishes and   
defining $\rho=\sum_{i=1}^{k}m_{\left(i\right)}n_{\left(i\right)}$,  Eqs.  (\ref{eq:4bcont}) and (\ref{eq:4-gradd}) reduce to
\begin{equation}
\frac{\partial}{\partial t}\rho+\nabla\cdot\left(\rho\vec{u}\right)=-\sum_{i=1}^{k}\nabla\cdot\vec{\mathcal{J}_{\left(i\right)}},
\end{equation}
\begin{equation}
\frac{\partial}{\partial t}\left(\rho\vec{u}\right)+\nabla\cdot\left(\overleftrightarrow{T}+\rho\vec{u}\vec{u}\right)=0,
\label{eq:4barinore}
\end{equation}
where $\overleftrightarrow{T}=\sum_{i=1}^{k}m_{\left(i\right)}n_{\left(i\right)}\left\langle \vec{C}_{\left(i\right)}\vec{C}_{\left(i\right)}\right\rangle $ is the usual stress tensor. In the next section a similar procedure is applied to the special relativistic case. The reader can notice that Eq. (\ref{eq:dif2}) also implies electric charge conservation  since
\begin{equation}
\frac{\partial }{\partial t}\left(\sum_{i=1}^{k}n_{(i)}e_{(i)}\right)+\nabla\cdot\left[\sum_{i=1}^{k}\left(n_{(i)}e_{(i)}\vec{u}\right)\right] =0.\label{eq:char}
\end{equation}In Eq. (\ref{eq:char}) $e_{(i)}$ represents  the charge corresponding to species $i$, which is also a collision invariant.Two possible generalizations of Eq. (\ref{eq:4-barinorel}) will arise, one will correspond to  the particle (Eckart) frame  and the other one to the energy  (Landau-Lifshitz) frame.

\section{The energy-momentum tensor for the relativistic binary mixture}
\noindent The Boltzmann equation for a mixture in the relativistic case reads  \cite{cercignani2002relativistic, jap, early} which reads:
\begin{equation}
V_{\left(i\right)}^{\mu}f_{\left(i\right),\mu}=\sum_{j}^ {k}
\mathsf{J}\left(f_{\left(j\right)}f_{\left(i\right)}\right),\quad i=1,2,\ldots,k.
\label{eq:4-Bolt2-1}
\end{equation}
where  $f_{\left(i\right)}$ is the distribution function for species $i$, $V_{\left(i\right)}^{\mu} $ is the single particle four-velocity given by $V_{(i)}^\mu=\left[\gamma_{(v,i)} v_{(i)}^\ell, c\gamma_{(v,i)}\right]$  and  $\mathsf{J}\left(f_{\left(j\right)}f_{\left(i\right)}\right)$ is the collision kernel. The Lorentz factor corresponds to $\gamma_{(v,i)}=1/\sqrt{1-\frac{v_{(i)}^2}{c^2}}$; in this definition $v_{(i)}^\ell $ is the particle three velocity. Greek indices will run form one to four, while latin indices will run from one to three. Eq. (\ref{eq:4-Bolt2-1}) can be generalized in order to include reactions which are relevant in high energy scenarios \cite{early}. Nevertheless, for the purpose of the present work we shall restrict to collisional interactions  assuming the absence of particle creation/annihilation processes. \\
In local equilibrium, the relativistic counterpart of Eq. (\ref{eq:Maxwell}) for species $i$ is the well-known J\"uttner function \cite{cercignani2002relativistic}:

\begin{equation}
f^{(0)}_{[J] (i)}=\frac{n_{(i)}}{4 \pi c^3 z_{(i)} \mathcal{K}_{2}\left(\frac{1}{z_{(i)}}\right)}\exp\left(-\frac{\gamma_{(v,i)}}{z_{(i)}}\right)
\label{eq:Juttner}
\end{equation}
where 
\begin{equation}
n_{(i)}=\int\gamma_{(v,i)}f^{(0)}_{[J](i)}d^{*}V,
\label{eq:n}
\end{equation} 
is the particle number density of species $i$, $c$ is the speed of light, $z_{(i)}=\frac{k_{B} T}{m_{(i)} c^2}$ corresponds to the relativistic parameter measuring the ratio between thermal energy and rest mass energy for each species, and $\mathcal{K}_{2}\left(\frac{1}{z_{(i)}}\right)$ is the modified Bessel function of the second kind.\\ 
The invariant measure $d^{*}V_{\left(i\right)}$
that appears in Eq. (\ref{eq:n}) corresponds to $\gamma_{\left(v,i\right)}^{5}\frac{dv_{\left(i\right)}^{3}}{v_{\left(i\right)}^{\left(4\right)}}$.
It is shown in Appendix B of Ref. \cite{garcia2011heat} that $d^{*}V_{\left(i\right)}=4\pi c^{3}\sqrt{\gamma_{\left(v,i\right)}^{2}-1}d\gamma_{\left(v,i\right)}$. This particular form of the invariant measure has been applied in
the framework of relativistic thermodynamics by several authors.\\
In order to establish the balance equations, we multiply Eq.  (\ref{eq:4-Bolt2-1}) by the collision  invariants $\left[m_{(i)},m_{(i)} V_{(i)}^\nu,e_{(i)}\right]$, obtaining upon integration the   particle flux:
\begin{equation}
J_{\left(i\right)}^{\mu }=m_{(i)}\int V_{\left(i\right)}^{\mu} f_{\left(i\right)}d^{*}V_{\left(i\right)},
\label{eq:4b}
\end{equation}	 
the energy-momentum tensor:

\begin{equation}
T_{\left(i\right)}^{\mu\nu}=m_{(i)}\int V_{\left(i\right)}^{\mu}V_{\left(i\right)}^{\nu}f_{\left(i\right)}d^{*}V_{\left(i\right)},
\label{eq:4-tensorere}
\end{equation}
and the electric current density: 

\begin{equation}
N_{(i)}^\mu=e_{(i)}\int V^\mu_{(i)} f_{(i)}d^*V_{(i)}.
\end{equation}
The invariant velocity element included in Eqs.(\ref{eq:4b}) and (\ref{eq:4-tensorere}) has been discussed by many authors since the original works by J\"uttner, and a brief calculation leading to its explicit form is included in the appendix of reference \cite{garcia2012microscopic}. The total balance equations are obtained summing over all species, leading to the expressions:
\begin{equation}
J^{\mu}_{,\mu}=0,\quad T^{\mu\nu}_{,\nu}=0.
\label{eq:4-bal}
\end{equation} 
Ordinary derivatives are used in Eqs.  (\ref{eq:4-bal})  since no  space-time curvature effects are relevant in this case.  
We now introduce the special relativistic generalization of the Galilean transformation of velocities (\ref{eq:4-vel}):

\begin{equation}
V_{\left(i\right)}^{\mu}=\mathcal{L}_{\alpha}^{\mu}K_{\left(i\right)}^{\alpha}=\gamma_{(K,i)}U^{\mu}+R_{\alpha}^{\mu}K_{\left(i\right)}^{\alpha}
\label{eq:veleck}
\end{equation}
where $\mathcal{L}_{\alpha}^{\nu}$ is the Lorentz transformation, $K_{\left(i\right)}^{\alpha} $ the chaotic four-velocity of an individual particle in the mixture, $\gamma_{(K,i)}$ is the Lorentz factor associated to the chaotic velocity of an individual particle of a given species, and $R_{\alpha}^{\mu}$ is the product of the Lorentz boost and the spatial projector $h_{\alpha}^{\beta}=\delta_{\alpha}^{\beta}+\frac{1}{c^2} U_{\alpha} U^{\beta}$. Equation (\ref{eq:veleck}) can be established using the fact that $V^\mu$ can be decomposed in terms of a  parallel component to the hydrodynamic velocity and another  one orthogonal to it. Thus $V^\mu_{(i)}=\gamma_{(v,i)} U^\mu+h^\mu_\nu V^\alpha_{(i)}$. Now introducing the chaotic velocity $K^\alpha_{(i)}$ in the right hand side of this equation one can easily obtain $V^\mu=\gamma_{(v,i)}U^\mu+h^\mu_\lambda L^\lambda_\alpha K^\alpha_{(i)}$, so that $R^\mu_\alpha=h^\mu_\lambda L^\lambda_\alpha$  \cite{garcia2012microscopic}. As it will be seen in the next subsections, the use of Eq. (\ref{eq:veleck}) leads to the special relativistic generalizations of equations (\ref{eq:4cont}) and (\ref{eq:4-gradd}).\\
The velocity included in Eq. (\ref{eq:veleck}) must refer to a flow velocity rather than to a reference system velocity. This concept is rather subtle and has been thoroughly discussed in references \cite{Grmale,V_n_2017}. In our case this quantity will refer to the hydrodynamic flows in the so-called particle and energy  frames, which will be discussed for multicomponent mixtures in the following subsections.\\
Notice that the total charge flux for the case of a plasma consisting of particles and antiparticles vanish since \\

\begin{equation}
N^{\mu}=\left(\sum_{i=1}^{k}e_{\left(i\right)}n_{\left(i\right)}\right)U^{\mu} +R_{\lambda}^{\mu}\sum_{i=1}^{k}e_{\left(i\right)}\int K_{\left(i\right)}^{\lambda}f_{\left(i\right)}d^{*}K_{\left(i\right)}.
\label{eq:27i}
\end{equation}
The first term corresponds to the local charge density which would vanish due to the presence of particle-antiparticle pairs. The second term would also vanish since each particle-antiparticle terms posses identical statistical properties and equal charges with opposite signs.  
\subsection{Energy-momentum tensor and center of mass velocity in the Eckart frame}

The particle balance equation for multicomponent mixtures is given by
\begin{equation}
J_{,\mu}^{\mu}=\sum_{i=1}^{k}m_{\left(i\right)}J_{\left(i\right),\mu}^{\mu}=0,
\end{equation}
where
\begin{equation}
\sum_{i=1}^{k}m_{\left(i\right)}J_{\left(i\right),\mu}^{\mu}=\sum_{i=1}^{k}m_{\left(i\right)}\int f_{\left(i\right)}V_{\left(i\right)}^{\mu}d^{*}V_{\left(i\right)}.
\label{eq:4part}
\end{equation}
Introducing Eq. (\ref{eq:veleck}) into Eq.  (\ref{eq:4part}), and taking into account only the spatial components of the balance equation  we immediately obtain

\begin{equation}
J^{\ell}=\sum_{i=1}^{k}m_{\left(i\right)}\int f_{\left(i\right)}\left(\gamma_{\left(K,i\right)}U^{\ell}+R_{\alpha}^{\ell}K_{\left(i\right)}^{\alpha}\right)d^{*}K_{\left(i\right)}.
\label{eq:4npart}
\end{equation}
We now define the statistical average in Eckart's frame as:

\begin{equation}
\left\langle \psi_{(i)} \right\rangle_{Eck}=\frac{1}{n_{(i)}}\int \psi_{(i)} f_{(i)} d^* V_{(i)}.
\end{equation}
The hydrodynamic velocity for species $i$ is given by

\begin{equation}
U^\ell_{(i)}=\left\langle V^\ell_{(i)}\right\rangle_{Eck}.
\end{equation}
Eq. (\ref{eq:4npart})  in this notation can be rewritten as:

\begin{equation}
J^{\ell}=U^{\ell}\left(\sum_{i=1}^{k}m_{\left(i\right)}n_{\left(i\right)}\right)+R_{\alpha}^{\ell}\left[\sum_{i=1}^{k}m_{\left(i\right)}n_{\left(i\right)}\left\langle K_{\left(i\right)}^{\alpha}\right\rangle _{Eck}\right].
\label{eq:DifEck}
\end{equation}
The last term in the right hand side of Eq. (\ref{eq:DifEck}) is the relativistic counterpart of the diffusion fluxes contained in Eq. (\ref{eq:dif2}). In Eckart's frame, these contributions are required to vanish, so that:
\begin{equation}
\sum_{i=1}^{k}m_{\left(i\right)}n_{\left(i\right)}\left\langle K_{\left(i\right)}^{\alpha}\right\rangle _{Eck}=0.
\label{eq:mk}
\end{equation}

\noindent Eq. (\ref{eq:mk}),  can be rewritten using  Eq. (\ref{eq:veleck}), thus obtaining:
\begin{equation}
-U^{\ell}\left(\sum_{i=1}^{k}m_{\left(i\right)}n_{\left(i\right)}\left\langle \gamma_{\left(v,i\right)}\right\rangle _{Eck}\right)+R_{\mu}^{\ell}\left(\sum_{i=1}^{k}m_{\left(i\right)}n_{\left(i\right)}\left\langle V_{\left(i\right)}^{\mu}\right\rangle _{Eck}\right)=0,
\label{eq:Lan}
\end{equation}
and the center  of mass velocity in the Eckart frame then reads:
\begin{equation}
U_{\left[Eck\right]}^{\ell}=\frac{R_{\mu}^{\ell}\left(\sum_{i=1}^{k}m_{\left(i\right)}n_{\left(i\right)}U_{\left(i\right)}^{\mu}\right)}{\sum_{i=1}^{k}m_{\left(i\right)}n_{\left(i\right)}\left\langle \gamma_{\left(v,i\right)}\right\rangle _{Eck}}=\frac{R_{\mu}^{\ell}\left(\sum_{i=1}^{k}m_{\left(i\right)}n_{\left(i\right)}U_{\left(i\right)}^{\mu}\right)}{\sum_{i=1}^{k}m_{\left(i\right)}n_{\left(i\right)}}
\label{eq:bariRelEck}
\end{equation}
Eq. (\ref{eq:bariRelEck}) is one possible relativistic analog of Eq. (\ref{eq:4-barinorel}). With this description the particle balance equations posses the same structure as in the non-relativistic formalism; nevertheless, relativistic effects from the heat flux arise in the spatial components of the  linear momentum balance equations.
The spatial components of the  energy-momentum tensor read:

\begin{equation}
T^{ab}=\sum_{i=1}^{k}m_{\left(i\right)}\int V_{\left(i\right)}^{a}V_{\left(i\right)}^{b}f_{\left(i\right)}d^{*}V_{\left(i\right)}
\label{eq:4T}
\end{equation}
Substituting Eq.  (\ref{eq:veleck}) in Eq. (\ref{eq:4T}) the energy-momentum tensor can be re-expressed as:
\begin{align}
T^{ab} & =U^{a}U^{b}\left(\sum_{i=1}^{k}m_{\left(i\right)}\int\gamma_{\left(K,i\right)}^{2}f_{\left(i\right)}d^{*}K_{\left(i\right)}\right)+R_{\alpha}^{a}R_{\beta}^{b}\left(\sum_{i=1}^{k}m_{\left(i\right)}n_{\left(i\right)}\left\langle K_{\left(i\right)}^{\alpha}K_{\left(i\right)}^{\beta}\right\rangle \right)\nonumber \\
& +\frac{U^{a}R_{\beta}^{b}}{c^{2}}\left(c^{2}\sum_{i=1}^{k}m_{\left(i\right)}\int\gamma_{\left(K,i\right)}K_{\left(i\right)}^{\beta}f_{\left(i\right)}d^{*}K_{\left(i\right)}\right)+\frac{U^{b}R_{\alpha}^{a}}{c^{2}}\left(c^{2}\sum_{i=1}^{k}m_{\left(i\right)}\int\gamma_{\left(K,i\right)}K_{\left(i\right)}^{\alpha}f_{\left(i\right)}d^{*}K_{\left(i\right)}\right)
\label{eq:4Tensorcomp}
\end{align}
The use of expression  (\ref{eq:mk}) leads, after straightforward algebraic manipulations to the Eckart's form of the spatial components of the energy momentum tensor: 
\begin{equation}
T^{a b}=\tilde{\rho}U^a U^b +\tau^{ab}+\frac{1}{c^2}U^a q^b+\frac{1}{c^2}U^b q^a,
\end{equation}
where:
\begin{equation}
\tilde{\rho}=\sum_{i=1}^{k}m_{\left(i\right)}\int\gamma_{\left(K,i\right)}^{2}f^{(0)}_{[J]\left(i\right)}d^{*}K_{\left(i\right)}
\end{equation}
is the mass-energy density which is defined in the local equilibrium state,
\begin{equation}
\tau^{ab}=R_{\alpha}^{a}R_{\beta}^{b}\left(\sum_{i=1}^{k}m_{\left(i\right)}n_{\left(i\right)}\left\langle K_{\left(i\right)}^{\alpha}K_{\left(i\right)}^{\beta}\right\rangle \right)
\end{equation}
is the stress tensor, and 
\begin{equation}
q^{a}=R_{\alpha}^{a}\left(c^{2}\sum_{i=1}^{k}m_{\left(i\right)}\int\gamma_{\left(K,i\right)}K_{\left(i\right)}^{\alpha}f_{\left(i\right)}d^{*}K_{\left(i\right)}\right)	
\end{equation}
is the heat flux. The inertial properties of heat, implied  by this last expression have been thoroughly discussed in Ref. \cite{Smerlak}.

In order to address the main non-equilibrium properties of the mixture we shall establish the corresponding expression for the entropy production of the system. For the sake of simplicity we will use the model equation first proposed by Marle, using a single collision time, for the treatment in Eckart's frame \cite{cercignani2002relativistic,Marle}  and the Anderson and Witting approach to the BGK formalism for the Landau-Lifshitz case \cite{struchtrup,Anderson}.\\	
The entropy balance equation for the mixture in Eckart's case is obtained upon multiplication of both sides of Eq. (\ref{eq:4-Bolt2-1}) by $k_B \ln \left(f_{\left(i\right)}\right)$ 
and integration with respect to the invariant measure $d^* V_{(i)}$. The corresponding result reads \cite{domi2017}:
\begin{equation}
\frac{\partial S^\nu}{\partial x^\nu}=\sigma
\label{eq:46}
\end{equation}
where $S^\nu$ is the entropy four-flux and $\sigma$ is the entropy production. These quantities are given by
\begin{equation}
S^\mu=-k_B \sum_{i=1}^{k}\int V_{(i)}^\mu f_{(i)}\ln\left( f_{(i)}\right)d^{*}V_{(i)}
\label{eq:entropp}
\end{equation}
and
\begin{equation}
\sigma=\frac{k_B}{\tau} \sum_{i=1}^{k}\int \left(f_{(i)}-f^{(0)}_{[J](i)}\right)\ln \left(f_{(i)}\right) d^{*}V_{(i)}
\label{eq:entropp1}
\end{equation}
where $\sum_{j=1}^{k}\mathsf{J}(f_{(i)}f_{(j)})$ is identified with $-\frac{f_{(i)}-f^{(0)}_{[J](i)}}{\tau}$ in Marle's model. The next step in order to obtain the entropy production is to introduce the Chapman-Enskog's hypothesis for the distribution function, that is
\begin{equation}
f_{(i)}=f_{[J](i)}^{(0)}\left(1+\varphi_{(i)}+\ldots\right)
\label{eq:49}
\end{equation}
where $f^{(0)}_{[J](i)}\varphi_{(i)}$ is a first order in the gradients correction to the local equilibrium distribution function $f_{[J](i)}^{(0)}$. A direct calculation  in the Navier-Stokes regime and using the fact that $\ln(1+\varphi_{(i)})\sim \varphi_{(i)}$ yields:
\begin{equation}
\sigma=\frac{k_B}{\tau}\sum_{i=1}^{k}\int f_{[J](i)}^{(0)}\varphi^2_{(i)}d^{*}V_{(i)},
\label{eq:entroleads}
\end{equation}
so that $\sigma$ is a positive semidefinite quantity. The explicit form of the entropy production in terms of the thermodynamic forces  is obtained through the expression 
\begin{equation}
\sigma^{(1)}\simeq-k_B\sum_{i=1}^{k}\int V^\alpha_{(i)}f^{(0)}_{(i),\alpha}\varphi_{(i)}d^{*}V_{(i)}
\end{equation}
where 
\begin{eqnarray}
V^\alpha_{(i)}f^{(0)}_{[J](i),\alpha}=&f^{(0)}_{[J](i)}\gamma_{(K,i)}h^\mu_\nu k^\nu_{(i)}\left\{\left(1-\frac{\gamma_{(K,i)}}{\mathcal{G}\left(\frac{1}{z_{(i)}}\right)}\right)\frac{n_{(i),\mu}}{n_{(i)}}+
\left(1-\frac{\gamma_{(K,i)}}{z_{(i)}}-\frac{\gamma_{(K,i)}}{\mathcal{G}\left(\frac{1}{z_{(i)}}\right)}-\frac{\mathcal{G}\left(\frac{1}{z_{(i)}}\right)}{z_{(i)}}\right)\frac{T_{,\mu}}{T}\right\} \label{eq:50}\\
&\left.
+\frac{1}{z_{(i)}c^{2}}\left[\gamma_{\left(K,i\right)}R_{\beta}^{\alpha}K_{\alpha\left(i\right)}\frac{h^{\beta\lambda}p_{,\lambda}}{\tilde{\rho}}+R_{\lambda}^{\alpha}R_{\beta}^{\mu}K_{\mu\left(i\right)}K_{\left(i\right)}^{\lambda}U_{\left(i\right),\alpha}^{\beta}\right]
\right.\nonumber
\end{eqnarray}
In Eq. (\ref{eq:50}),   $\mathcal{G}\left(\frac{1}{z_{(i)}}\right)$ is the ratio of $\mathcal{K}_{2}\left(\frac{1}{z_{(i)}}\right)$ and $\mathcal{K}_{3}\left(\frac{1}{z_{(i)}}\right)$, where $\mathcal{K}_{n}\left(\frac{1}{z_{(i)}}\right)$ corresponds to the modified Bessel function of the $n$-th kind. The establishment of fluxes requires a careful election of the representation of the thermodynamic forces in order to satisfy Onsager's reciprocity relations. The corresponding analysis for  Eckart's frame is discussed elsewhere \cite{jap,moratto} and is beyond  the scope of the present work. This subject will be part of a thorough analysis in the near future. 

\subsection{Energy-momentum tensor and center of mass velocity in the Landau and Lifshitz frame}

In order to grasp   the role of the heat flux in the Landau-Lifshitz's frame, we start considering Eq. (\ref{eq:Lan}),  introducing a slightly different definition of statistical average, namely

\begin{equation}
\left\langle \psi_{\left(i\right)}^{a}\right\rangle _{\left[\mathcal{LL}\right]}=\frac{1}{n_{(i)}}\int\gamma_{\left(K,i\right)}\psi_{\left(i\right)}^{a}f_{(i)}d^{*}V_{\left(i\right)},
\label{eq:averLL}
\end{equation}
The introduction of the new notation and using Eq. (\ref{eq:veleck}) in order to rewrite Eq. (\ref{eq:4T}) leads to

\begin{align}
T^{k\ell} & =U^{k}U^{\ell}\left({\sum_{i=1}^{k}m_{\left(i\right)}\int\gamma_{\left(K,i\right)}^{2}f_{\left(i\right)}d^{*}K_{\left(i\right)}}\right)+R_{\alpha}^{k}R_{\beta}^{\ell}\left({\sum_{i=1}^{k}m_{\left(i\right)}\int K_{\left(i\right)}^{\alpha}K_{\left(i\right)}^{\beta}f_{\left(i\right)}d^{*}K_{\left(i\right)}}\right)\nonumber \\
& +U^{k}R_{\beta}^{\ell}\left({\sum_{i=1}^{k}m_{\left(i\right)}n_{\left(i\right)}\left\langle K_{\left(i\right)}^{\beta}\right\rangle }\right)+U^{\ell}R_{\alpha}^{k}\left({\sum_{i=1}^{k}m_{\left(i\right)}n_{\left(i\right)}\left\langle K_{\left(i\right)}^{\alpha}\right\rangle }\right).
\end{align}
We now proceed in a similar fashion as in the establishment of Eq. (\ref{eq:4-barinorel}) form Eq. (\ref{eq:4-gradd}). The condition necessary for the cross-terms in the momentum balance to vanish now read:
\begin{equation}
{\sum_{i=1}^{k}m_{\left(i\right)}n_{\left(i\right)}\left\langle K_{\left(i\right)}^{\ell}\right\rangle _{\left[\mathcal{LL}\right]}=0.}
\label{eq:4-sumchaorel}	
\end{equation}
so that the center of mass velocity in the Landau-Lifshitz frame takes the form
\begin{equation}
U_{\left[\mathcal{LL}\right]}^{\ell}=\frac{R_{\alpha}^{\ell}\left({\sum_{i=1}^{k}m_{\left(i\right)}n_{\left(i\right)}U_{\left(i\right)}^{\alpha}}\right)}{{\sum_{i=1}^{k}\left(m_{\left(i\right)}\int\gamma_{\left(K,i\right)}^{2}f_{\left(i\right)}d^{*}K_{\left(i\right)}\right)}}.
\end{equation}
The type of integral present in the denominator of Eq. (\ref{eq:internal}) is well-known and corresponds to the internal energy density of each species \cite{cercignani2002relativistic,garcia2012microscopic}, namely
\begin{equation}
\varepsilon_{(i)}=m_{(i)}c^2\int \gamma^2_{(K,i)}f^{(0)}_{[J](i)}d^{*}K_{(i)}=n_{(i)} m_{(i)} c^2\left(3z_{(i)}+\frac{\mathcal{K}_{1}\left(\frac{1}{z_{(i)}}\right)}{\mathcal{K}_{2}\left(\frac{1}{z_{(i)}}\right)}\right).
\label{eq:internal}
\end{equation}	
On the other hand, new fluxes  are present in the particle balance equation, which now reads:
\begin{equation}
{J^{a}=U_{\left[\mathcal{LL}\right]}^{a}\sum_{i=1}^{k}n_{\left(i\right)}m_{\left(i\right)}+\sum_{i=1}^{k}\mathcal{J}_{\left(i\right)}^{a}},
\end{equation}
The sum of the diffusion fluxes 
\begin{equation}
{\mathcal{J}_{\left(i\right)}^{a}}=R_{\alpha}^{a}{\sum_{i=1}^{k}\left(m_{\left(i\right)}\int K_{\left(i\right)}^{\alpha}f_{\left(i\right)}d^{*}K_{\left(i\right)}\right)}.
\label{eq:fluxdi}
\end{equation}
Out of equilibrium, the sum of the diffusion fluxes \textit{does not vanish} in contrast with the non-relativistic case. The specific form of this strictly relativistic effects  depends on the non-equilibrium distribution function derivable from approximate solutions of the Boltzmann equation (\ref{eq:4-Bolt2-1}).

The reader must notice that Eq. (\ref{eq:fluxdi}) is not in conflict with charge conservation in the case  of a system consisting in particle-antiparticle pairs.  This can be seen from Eq. (\ref{eq:27i}) which admits opposite signs that lead to the cancellation of the total flux   in the presence of symmetry. This is not the case for the mass flux (\ref{eq:fluxdi}) since all masses are non-negative.\\
The entropy production of the multicomponent system in the Landau-Lifshitz frame can be established following the same  procedure presented in the previous section. In this case, the Boltzmann equation in the BGK approximation takes the form \cite{struchtrup,Anderson}:
\begin{equation}
V_{\left(i\right)}^{\mu}f_{\left(i\right),\mu}=-\gamma_{(K,i)}\frac{f_{(i)}-f^{(0)}_{(i)}}{\tau}
\label{eq:4-Bolt2-3}
\end{equation}
Eqs. (\ref{eq:46}), (\ref{eq:entropp}) and (\ref{eq:49})  remain invariant in both frames, but in the Landau-Lifshitz case Eqs.  (\ref{eq:entropp1}) and (\ref{eq:entroleads}) become:
\begin{equation}
\sigma_{[\mathcal{LL}]}=\frac{k_B}{\tau}\sum_{i=1}^{k}\int \gamma_{(K,i)}\left(f_{(i)}-f_{[J](i)}^{(0)}\right)\ln \left(f_{(i)}\right)d^{*}V_{(i)}
\end{equation} 
so that
\begin{equation}
\sigma_{[\mathcal{LL}]}=\frac{k_B}{\tau}\sum_{i=1}^{k}\int \gamma_{(K,i)} f^{(0)}_{[J](i)}\varphi^2_{(i)}d^{*}V_{(i)}
\label{eq:ent}
\end{equation}
The entropy production given in Eq. (\ref{eq:ent}) is positive semidefinite and can be expressed in  terms of the first order in the gradients  thermodynamic forces as:
\begin{equation}
\sigma_{[\mathcal{LL}]}^{(1)}\simeq-k_B\sum_{i=1}^{k}\int \gamma_{(K,i)}V_{(i)}^\alpha f^{(0)}_{(i),\alpha}\varphi_{(i)}d^{*}V_{(i)}
\label{eq:62}
\end{equation}
where $V_{(i)}^\alpha f^{(0)}_{[J](i),\alpha}$ is still given by Eq. (\ref{eq:4-Bolt2-3}). It is important to emphasize that the $\gamma_{(K,i)}$ factor in the integral at the right hand side of Eq. (\ref{eq:62}) leads in the Landau-Lifshitz case to non-vanishing contributions to the entropy production corresponding to the scalar product of the particle flux and its corresponding thermodynamic forces. This feature is not present in Eckart's frame (see Eq. (35) in Appendix B of Ref. \cite{domi2017}).   In Table \ref{table:1} we compare both descriptions of relativistic kinetic theory in terms of the aforementioned operational definitions of the center of mass velocity of the relativistic mixture.

\section{Final Remarks}
The first formulations of relativistic irreversible thermodynamics were firstly proposed by purely phenomenological arguments  \cite{eckart1940thermodynamics, Landau}. Pathological features of the early formalism were later identified suggesting the need of extended theories \cite{Jou1996}. On the other hand, a thorough revision of the microscopic foundations of the transport theory of dissipative relativistic fluid has led to stable sets of equations within the Chapman-Enskog method and the use of linear constitutive equations \cite{GARCIAPInest}.

It has been shown that the structure of the energy-momentum tensor in the relativistic kinetic theory of multicomponent mixtures depends on the operational definition of the center of mass velocity of the system. We believe that this physical insight of Eckart's and Landau-Lifshitz's frames will be useful for explicit calculations regarding transport properties of relativistic multicomponent systems. Immediate work using the present approach include new studies regarding  the relativistic Onsager's reciprocity relations \cite{moratto}, and the analysis of the entropy production for a relativistic multicomponent mixture to first order in the gradients in the Landau-Lifshitz frame. Other interesting future work corresponds to the study of relativistic multicomponent systems in the realm of Kaluza's magnetohydrodynamics \cite{sagaceta2016statistical}, and a thorough revision of the generic instabilities of linear perturbations in the case of relativistic multicomponent systems \cite{cercignani2002relativistic}.

On a future paper we will discuss the  time-like components of the balance equations  in order to establish a complete description of the system. We naturally expect that the methods used in Ref. \cite{SandovalVillalbazoPhisicaA} applied to the corresponding linearized system will show stability properties for the multicomponent system in the energy frame.



\begin{thebibliography}{10}
	
	\bibitem{eckart1940thermodynamics}  C. Eckart, The thermodynamics of irreversible processes. III. Relativistic theory of the simple fluid, \emph{Physical Review} \textbf{58} (1940), 919-924.	
	
	\bibitem{Landau}L. Landau and E. Lifshitz, \emph{Fluid Mechanics}, vol. 6, Elsevier Science, 2013.
	
	\bibitem{Israel} W. Israel, Relativistic kinetic theory of a simple
	gas,  \emph{Journal of Mathematical Physics} \textbf{4} (1963), pp.
	1163-1181.
	
	\bibitem{struchtrup} H. Struchtrup, Projected moments in relativistic kinetic
	theory, \emph{Physica A: Statistical Mechanics and its Applications}, \textbf{253} (1998), pp. 555-595.
	
	\bibitem{cercignani2002relativistic} C. Cercignani and G. M. Kremer,  \emph{The Relativistic Boltzmann Equation: Theory and Applications}, pp. 31-63, Springer, 2002.
	
	\bibitem{garcia2012microscopic} A.L. García-Perciante, A. Sandoval-Villalbazo, and L. García-Colín, On the microscopic nature of dissipative effects in special relativistic
	kinetic theory, \emph{Journal of Non-Equilibrium Thermodynamics} \textbf{37} (2012), pp. 43-61.
	
	\bibitem{Hiscock} W. A. Hiscock and L. Lindblom, Generic instabilities
	in first order dissipative relativistic fluid theories, \emph{Phys. Rev. D} \textbf{31} (1985), pp. 725-733.
	
	\bibitem{PeterVan} P. Ván, Generic stability of dissipative non-relativistic and relativistic fluids, \emph{Journal of Statistical Mechanics: Theory and Experiment} \textbf{2009} (2009), p. P0254.
	
	\bibitem{Koide} S. Pu, T. Koide, and D. H. Rischke, Does stability
	of relativistic dissipative fluid dynamics imply causality?, \emph{Phys. Rev. D} \textbf{81} (2010), p. 114039.
	
	\bibitem{Smerlak} M. Smerlak, On the inertia of heat, \emph{The European Physical Journal Plus} \textbf{127} (2012), p 72.
	
	\bibitem{SandovalVillalbazoPhisicaA} A. Sandoval-Villalbazo, A. L. García-Perciante, and L. S. García-Colín, Relativistic transport theory for simple fluids at	first order in the gradients: A stable picture, \emph{Physica A}  \textbf{388} (2009), p. 3765.
	
	\bibitem{muller} I. Müller, Toward relativistic thermodynamics, \emph{Archive for Rational Mechanics and Analysis}  \textbf{34} (1969), p. 259-282.
	
	\bibitem{israelextended} W. Israel, Nonstationary irreversible thermodynamics: A casual relativistic theory,  \emph{Annals of Physics}  \textbf{100} (1976), p. 310-331.
	
	\bibitem{stewart} M. Stewart, On transient relativistic thermodynamics and kinetic theory,  \emph{Proceedings of the Royal Society of London. Series A, Mathematical and Physical Sciences}  \textbf{357} (1977), p. 59-75.
	
	\bibitem{Tsumura} K. Tsumura and T. Kunihiro, Uniqueness of Landau-Lifshitz energy frame in relativistic dissipative hydrodynamics  \emph{Physical Review E},  \textbf{87} (2013), p. 053008.
	
	\bibitem{GARCIAPInest} A. L. García-Perciante and A. Sandoval-Villalbazo, Remarks on relativistic kinetic theory to first order in the gradients, \emph{Journal of Non-Newtonian Fluid Mechanics} \textbf{165} (2010), pp. 1024-1028.
	
	\bibitem{Chapman1970} S. Chapman and T. Cowling, \emph{The mathematical theory of non-uniform Gases: An account of the kinetic theory of viscosity, thermal conduction and diffusion in gases}, Cambridge Mathematical Library, Cambridge University Press, 1970.
	
	\bibitem{jap} Y. Kikuchi, K. Tsumura and T. Kunihiro, Derivation of second-order relativistic hydrodynamics for reactive multicomponent systems,  \emph{Physical Review C}  \textbf{92} (2015), p. 064909.
	
	\bibitem{early} E. Kolb and M. Turner, \emph{The Early Universe}, Frontiers in physics, Addison-Wesley, 1990.
	
	\bibitem{garcia2011heat} A. L. García-Perciante and A. R. Méndez, Heat conduction in relativistic neutral gases revisited, \emph{General Relativity and Gravitation} \textbf{42} (2011), pp. 2257-2275.
	
	\bibitem{Grmale} P. Ván, M. Pavelka and M. Grmela, Extra mass flux in fluid mechanics, \emph{Journal of Non-Equilibrium Thermodynamics} \textbf{42} (2017), pp. 133-151.
	
	\bibitem{V_n_2017} P. Ván, Galilean relativistic fluid mechanics, \emph{Continuum Mechanics and Thermodynamics} \textbf{29} (2017), pp. 585-610.
	
	\bibitem{Marle} C. Marle, Sur l'établissement des équations de l'hydrodynamique des fluides relativistes dissipatifs. i. l'equation de Boltzmann relativiste, \emph{Anna es de l'l.H.P. Physique théorique} \textbf{10} (1969), pp. 67-126.
	
	\bibitem{Anderson} J. Anderson and H. Witting, A relativistic relaxation-time model for the Boltzmann equation, \emph{Physica} \textbf{74} (1974), pp. 466-488.
	
	\bibitem{domi2017} D. Brun-Battistini, A. Sandoval-Villalbazo and A. L. García-Perciante, Entropy production in simple special relativistic fluids, \emph{Journal of Non-Equilibrium Thermodynamics} \textbf{39} (2014), pp. 27-33.
	
	\bibitem{moratto} V. Moratto  A. L. García-Perciante, and L. S. García-Colín,  Validity of the Onsager relations in relativistic binary mixtures, \emph{Physical Review E, Statistical, nonlinear, and soft matter physics}, \textbf{84} (2011), p. 021132.
	
	\bibitem{Jou1996} D. Jou, J. Casas-Vazquez, and G. Lebon, \emph{Extended Irreversible Thermodynamics}, 	4th ed., Springer-Verlag Berlin Heidelberg, 2001.
	
	\bibitem{sagaceta2016statistical}	A. R. Sagaceta-Mejía and Sandoval-Villalbazo, On the
	statistical foundations of magnetohydrodynamics, \emph{American
		Institute of Physics Conference Series} \textbf{1786} (2016), p.040007.
	
\end{thebibliography}
\begin{table}
	\begin{tabular}{|c|c|c|}
		\hline 
		& Eckart & Landau\tabularnewline
		\hline 
		\hline 
		Statistical Average & $\left\langle \psi_{\left(i\right)}^{a}\right\rangle _{\left[Eck\right]}=\frac{1}{n_{\left(i\right)}}\int\psi_{\left(i\right)}^{a}f_{\left(i\right)}d^{*}K_{\left(i\right)}$ & $\left\langle \psi_{\left(i\right)}^{a}\right\rangle _{\left[\mathcal{LL}\right]}=\frac{1}{n_{\left(i\right)}}\int\gamma_{\left(K,i\right)}\psi_{\left(i\right)}^{a}f_{\left(i\right)}d^{*}K_{\left(i\right)}$\tabularnewline
		\hline 
		Center of mass velocity & $U^{\ell}=\frac{\mathcal{R}_{\mu}^{\ell}\left({\sum_{i=1}^{k}n_{(i)}m_{(i)}U_{(i)}^{\mu}}\right)}{{\sum_{i=1}^{k}n_{(i)}m_{(i)}}}$ & $U_{[\mathcal{LL}]}^{\ell}=\frac{\mathcal{R}_{\mu}^{\ell}\left({\sum_{i=1}^{k}n_{(i)}m_{(i)}U_{(i)}^{\mu}}\right)}{\frac{1}{c^{2}}\left({\sum_{i=1}^{k}n_{(i)}\varepsilon_{(i)}}\right)}$\tabularnewline
		\hline 
		Energy-momentum tensor & $T^{ab}=\tilde{\rho}U^{a}U^{b}+\tau^{ab}+\frac{1}{c^{2}}U^{a}q^{b}+\frac{1}{c^{2}}U^{b}q^{a}$ & $T^{ab}=\tilde{\rho}U_{[\mathcal{LL}]}^{a}U_{[\mathcal{LL}]}^{b}+\tau^{ab}$\tabularnewline
		\hline 
		Particle flux & 
		$
		J^{a}=\left({\sum_{i=1}^{k}n_{(i)}m_{(i)}}\right)U^{a}
		$
		& $J^{a}=\left({\sum_{i=1}^{k}n_{(i)}m_{(i)}}\right)U_{[\mathcal{LL}]}^{a}+{\sum_{i=1}^{k}\mathcal{J}_{\left(i\right)}^{a}}$\tabularnewline
		\hline 
	\end{tabular}
	\caption{Comparative table for  the relativistic kinetic theory of multicomponent mixtures in  Eckart's and  Landau-Lifshitz's frames.}
	\label{table:1}
\end{table}

\end{document}